\documentclass[12pt]{article}
\usepackage{amsmath}
\usepackage{graphicx}
\usepackage{natbib}
\newcommand{\bigCI}{\mathrel{\text{\scalebox{1.07}{$\perp\mkern-10mu\perp$}}}}
\usepackage{amsfonts}
\usepackage{amssymb}
\usepackage{graphicx}
\usepackage{times}
\usepackage{ragged2e}
\usepackage{adjustbox}
\usepackage{multirow}
\usepackage{colortbl}
\newtheorem{theorem}{Theorem}
\newtheorem{corollary}{Corollary}[theorem] %
\usepackage{caption}
\usepackage{subcaption}

\usepackage{booktabs,hyperref}
\usepackage{float,caption,hypcap}
\providecommand{\numberTblEq}[1]{\refstepcounter{tblEqCounter}\label{#1}\thetag{\thetblEqCounter}}

\justifying
\usepackage[T1]{fontenc} 

\usepackage{url} % not crucial - just used below for the URL 

\newcommand{\blind}{0}

\begin{document}
\newcounter{tblEqCounter} %create a counter

\def\spacingset#1{\renewcommand{\baselinestretch}%
{#1}\small\normalsize} \spacingset{1}

%%%%%%%%%%%%%%%%%%%%%%%%%%%%%%%%%%%%%%%%%%%%%%%%%%%%%%%%%%%%%%%%%%%%%%%%%%%%%%

\if0\blind
{
  \title{\bf A New Covariance Estimator for Sufficient Dimension Reduction in High-Dimensional and Undersized Sample Problems}
  \author{Kabir Opeyemi Olorede\thanks{
    Kabir Opeyemi Olorede is a PhD candidate in the Department of Statistics, University of Ilorin, Ilorin, Nigeria.}\hspace{.2cm}\\ Department of Statistics and Mathematical Sciences,\\ Kwara State University, Malete, Nigeria\\
    and \\
    Waheed Babatunde Yahya\thanks{Waheed Babatunde Yahya is a Professor of Biostatistics and Data Mining at University of Ilorin, Ilorin, Nigeria}\\Department of Statistics, University of Ilorin, Ilorin, Nigeria}
  \maketitle
}\fi

\if1\blind
{
  \bigskip
  \bigskip
  \bigskip
  \begin{center}
    {\LARGE\bf Title}
\end{center}
  \medskip
} \fi
\bigskip
\begin{abstract}
The application of standard sufficient dimension reduction methods for reducing the dimension space of predictors without losing regression information requires inverting the covariance matrix of the predictors. This has posed a number of challenges especially when analyzing high-dimensional data sets in which the number of predictors $\mathit{p}$ is much larger than number of samples $n,~(n\ll p)$. A new covariance estimator, called the \textit{Maximum Entropy Covariance} (MEC) that addresses loss of covariance information when similar covariance matrices are linearly combined using \textit{Maximum Entropy} (ME) principle is proposed in this work. By benefitting naturally from slicing or discretizing range of the response variable, y into \textit{H} non-overlapping categories, $\mathit{h_{1},\ldots ,h_{H}}$, MEC first combines covariance matrices arising from samples in each y slice $\mathit{h\in H}$  and then select the one that maximizes entropy under the principle of maximum uncertainty. The MEC estimator is then formed from convex mixture of such entropy-maximizing sample covariance $S_{\mbox{mec}}$  estimate and pooled sample covariance $\mathbf{S}_{\mathit{p}}$  estimate across the $\mathit{H}$ slices without requiring time-consuming covariance optimization procedures. MEC deals directly with singularity and instability of sample group covariance estimate in both regression and classification problems. The efficiency of the MEC estimator is studied with the existing sufficient dimension reduction methods such as \textit{Sliced Inverse Regression} (SIR) and \textit{Sliced Average Variance Estimator} (SAVE) as demonstrated on both classification and regression problems using real life Leukemia cancer data and customers' electricity load profiles from smart meter data sets respectively.
\end{abstract}
\noindent%
{\it Keywords:}  Slicing, Sufficient Dimension Reduction, Central Subspace, Loss of Covariance \\Information, Maximum Entropy Covariance.
\vfill 
\newpage
\spacingset{1.5} % DON'T change the spacing!
\section{Introduction}
\label{sec:intro}
With the recent surge of interest in addressing the potentially daunting statistical learning and pattern recognition problems presented by the \textquotedblleft curse of dimensionality\textquotedblright, \cite{Bellman1961} when analyzing high-dimensional data sets in which number of predictors, $\mathit{p}$ is much larger than the number of samples, \textit{n} $(\mathit{n}\ll \mathit{p})$, many dimension reduction techniques have been developed.  Some of these techniques are non-probabilistic and unsupervised (reduction is done only on predictor vector $\mathbf{X}$), such as classical Principal Component Analysis (PCA: \citeauthor{Jollife2002}, \citeyear{Jollife2002}); some are unsupervised and probabilistic, such as Factor Analysis (FA), Probabilistic Principal Component Analysis (PPCA: \citeauthor{TippingBishop1997} \citeyear{TippingBishop1997}; \citeauthor{TippingBishop1999}, \citeyear{TippingBishop1999}), Kernel Principal Component Analysis (KPCA); some are two-stage supervised and non-probabilistic, such as Principal Component Regression (PCR), Partial Least Squares Regression (PLSR: \citeauthor{Yahyaetal2017}, \citeyear{Yahyaetal2017}); while some are supervised probabilistic, such as Elastic Net (EN) regression models, PPCA plus Adaptive Elastic Net (AEN), regression models, to mention just a few.\\   
Considering a high-dimensional regression or classification problem involving a discrete or continuous univariate response $\mathbf{Y}$ and p-dimensional predictor vector $\mathbf{X}=(x_{1},\ldots ,x_{p})^{T}$ as often the case while analyzing a whole genome-wide SNP dataset \citep{LiYin2008}, microarray-based gene expression data~\citep{Yahyaulm2011, YahyaRosenberg2014} and electricity smart meter data in electricity consumption and billing profiling problems \citep{TureczekNielsen2017}. Replacing $\mathbf{X}$ by a lower dimensional function R($\mathbf{X}$) that captures most variance in $\mathbf{X}$ or most regression information of $\mathbf{Y}$ on $\mathbf{X}$ is called Dimension Reduction (DR). With motivations in visualization of data, mitigation of dimensionality issues in estimating the mean conditional function $E(\mathbf{Y}|\mathbf{X})$ and better prediction of future observations, dimension reduction methods have been successful in genomics literature (see for example, \citeauthor{Golub1999}, \citeyear{Golub1999}; \citeauthor{Dudoit2002}, \citeyear{Dudoit2002}; \citeauthor{BuraPfeiffer2003}, \citeyear{BuraPfeiffer2003}; and \citeauthor{Pamukcu2015}, \citeyear{Pamukcu2015}) and as an important step in the supervised principal components methods of \cite{Bair2006}.\\ 
Dimension reduction methods seek to organize the variations in the data in an interpretable way, according to the magnitudes of variations often based on arbitrary stopping rules.  However, finding a reduction R($\mathbf{X}$) of dimension $d < p$ that captures all regression information of $\mathbf{Y}$ on $\mathbf{X}$ is called sufficient dimension reduction \citep{Cook1994, Cook1996, Li1991}. Sufficient dimension reduction (SDR) theory \citep{Cook1998} has been developed to reduce the predictor dimension while preserving full regression information and without requiring a pre-specified parametric model for $\mathbf{Y}|\mathbf{X}$. Based on the notion of sufficiency, SDR differs from DR in that it organizes the variations in the predictor according to how much they can explain the response variables \citep{Li2018}.\\ 
Let $d < min(n,p)$ and let $\beta_{1}^{T} \mathbf{X},\ldots,\beta_{d}^{T} \mathbf{X} \in \mathbb{R}^{p}$ define smallest number of first few linear combinations of the stochastic covariate vector $\mathbf{X}$ so that 
\begin{equation}\label{eq:1}
\mathbf{Y}\bigCI \mathbf{X}|(\beta_{1}^{T} \mathbf{X},\ldots , \beta_{d}^{T} \mathbf{X}),
\end{equation}
where $\bigCI$ signifies statistical independence which implies that $\mathbf{Y}$ is independent of $\mathbf{X}$ given the $\mathit{d}$ linear combinations $\beta_{1}^{T} \mathbf{X},\ldots ,\beta_{d}^{T} \mathbf{X}$  of $\mathbf{X}$ by placing no restrictions on the regression in equation (\ref{eq:1}). If (\ref{eq:1}) is true, the linear combinations $\beta_{1}^{T} \mathbf{X},\ldots, \beta_{d}^{T} \mathbf{X}$  are called sufficient dimension reduction directions or sufficient predictors because they contain all the regression information that $\mathbf{X}$ has about $\mathbf{Y}$. Without loss of generality, we let $\beta_{1}^{T} \mathbf{X},\ldots,\beta_{d}^{T} \mathbf{X}$ be the $\beta_{j},~j=1,\ldots ,d$ columns of the $p\times d$ matrix $\mathbb{B}$ and replace the $p\times 1$ predictor vectors $\mathbf{X}$ by the sufficient predictors $\mathbb{B}^{T} \mathbf{X}$. A dimension reduction subspace $\mathbf{S}$ is defined as the subspace spanned by $\beta_{1}^{T} \mathbf{X},\ldots,\beta_{d}^{T} \mathbf{X}$  such that if $\mbox{span} (\mathbb{B})= S_{\mathbf{Y}|\mathbf{X}}$,  then $\mathbf{Y}\bigCI \mathbf{X}|\mathbb{B}^{T} \mathbf{X}$. The intersection of all such subspace $\mathbf{S}$, if itself satisfies the conditional independence, is called the central subspace \citep{Cook1994, Cook1996} with dimension  $d = \dim(S_{\mathbf{Y}|\mathbf{X}})$. Therefore, sufficient dimension reduction of the predictor vector $\mathbf{X}$ amounts to estimating a basis for the meta-parameter $\mathbf{S}_{\mathbf{Y}|\mathbf{X}}$ and its dimension $\mathit{d}$ \cite{Hilafu2017}. The Central Subspace (CS), $\mathbf{S}_{\mathbf{Y}|\mathbf{X}}$, is a well-defined, unique and parsimonious population parameter under some mild conditions including linearity condition and constant covariance condition \citep{Cook1996, YinLiCook2008} that satisfy (\ref{eq:1}).  Therefore, without loss of generality, sufficient dimension reduction of the predictor vector $\mathbf{X}$ amounts to estimating a basis for the meta-parameter $S_{\mathbf{Y}|\mathbf{X}}$ and its dimension $\mathit{d}$. \\
Numerous SDR methods have been proposed in the statistics literature since the seminar paper of ~\cite{Li1991} on Sliced Inverse Regression (SIR). Among these methods are the Sliced Average Variance Estimation (SAVE, ~\citeauthor{CookWeisberg1991}, \citeyear{CookWeisberg1991}), Principal Hessian Directions (PHD, ~\citeauthor{Li1992}, \citeyear{Li1992}), Directional Regression \citep{LiWang2007}, and the Inverse Regression Estimation (IRE, ~\citeauthor{CookNi2005}, \citeyear{CookNi2005}).\\
Many authors have proposed a plethora of methods to estimate the central subspace. Attractive computational properties of these methods as inverse conditional moments \citep{CookWeisberg1991, Li1991, CookYin2002} have led to their extensive use in diverse applications. A comprehensive list of references on vast literature of sufficient dimension reduction is provided in \cite{MaZhu2013b}. However, involvement of covariance matrix inversion in the basic step of these methods has plagued their success in applications where data sets contain high-dimensional predictors,$\mathit{p}$ and undersized samples, $\mathit{n}~(n\ll p)$ due to covariance matrix ill-conditioning or eigenvalue degeneracy which poses serious challenge to the computational tools.\\
There have been different proposals in the existing literature to circumvent this potentially daunting statistical problem. \cite{ChiaromonteMartinelli2002} proposed using a two-stage approach involving the use of Singular Value Decomposition (SVD) to reduce the predictor dimension at first stage and then apply SDR methods to the reduced $\mathit{d} < \mathit{n}$ principal components. The PCA involved in their two-stage problem for identification of predictive components is known to suffer several shortcomings. It does not work well in limited sample problems since the estimated covariance matrix becomes rank deficient \citep{NaikTsai2000, Lietal2007, Cooketal2007, ZhuZhu2009b, Zhuetal2010b, HilafuYin2017, Tanetal2018}. The Principal Components (PCs) are computed from the predictors alone and do not make apparent use of the response.  The PCs are not invariant or equivariant under full rank linear transformation of the predictors. Finally, the determination of optimal number of components to include as predictors in the second stage is heavily dependent on some arbitrary rules. As an alternative method which avoids covariance matrix inversion, \cite{Cooketal2007} proposed seeded sufficient dimension reduction methods based on the ideas from the partial least squares estimation \citep{Helland1990}.  They noted that their proposed methods cannot tackle contemporary large-$\mathit{p}$-small-$\mathit{n}$  regressions in which information accumulates as $\mathit{p}$ grows. \\ 
\cite{Zhong2005} proposed regularized sliced inverse regression by adding a product of a constant term and the identity matrix to the covariance matrix of the predictors. Their regularization strategy requires time consuming optimization. \cite{LiYin2008}   proposed a regularized least-squares formulation of sliced inverse regression by introduction of $\ell_{1}$ and $\ell_{2}$ norm penalties in a unified approach with the development of an alternating least-squares approach. This formulation also relies on time-consuming optimization and parameter tuning. \\
Based on intelligent partitioning of the predictors into smaller subsets of predictors and sequential reduction of these smaller subsets, \cite{YinHilafu2015} proposed a general sequential dimension framework that circumvents the curse of dimensionality issues in sufficient dimension reduction with sliced inverse regression. This sequential framework is computationally intensive and the differences between solutions from different partitions of the predictor vector may be difficult to quantify rigorously, though, an ensemble idea can provide a promising solution. \cite{Hilafu2017} proposed random sliced inverse regression and cluster-wise sliced inverse regression as computationally intensive methods that circumvent the covariance singularity issues in sliced inverse regression \citep{Li1991} based on the idea of random forest \citep{Breiman2001}.  Performance of this method depends heavily on the choice of bootstrap samples, and can only apply SIR to randomly selected candidate variables that are fewer than the sample size. \\
In this paper, a new covariance estimator called Maximum Entropy Covariance (MEC) estimator which effectively deals with the singularity and instability of sample covariance estimate in sufficient dimension reduction application has been develop. The MEC estimator is based on ideas from Maximum Entropy Covariance Selection (MECS) and Eigenvalue Stabilization of \cite{Thomaz2004}. \\
Other sections in this paper constitute the following: The Sliced Inverse Regression and the Sliced Average Variance Estimation methods are reviewed in section two. Also, reviews of loss of covariance information, the maximum entropy principle and the existing Maximum Entropy Covariance Selection (MECS) method are presented in section three. Section four presents the proposed Maximum Entropy Covariance estimator. The usual MLE covariance estimators in sufficient dimension reduction methods are replaced with MEC and they are applied to real life data sets involving undersized samples and high-dimensional predictors in Section five. Discussions of results and conclusion are presented in Section six.  
\section{Review of Sufficient Dimension Reduction Methods}
\label{sec:review}
As basic motivation for MEC proposal, a short review of sufficient dimension reduction methods including SIR \citep{Li1991} and SAVE \citep{CookWeisberg1991} is presented.
\subsection{The Sliced Inverse Regression (SIR)}
\label{subsec:SIR}
The core idea of SIR is based on the assumption that:  
\begin{quotation}
for any vector $b\in \mathbb{R}^{p}$,   $E(b^{T} \mathbf{X}|\mathbb{B}^{T} \mathbf{X})$ is a linear function of $\mathbb{B}^{T} \mathbf{X}$. 
\end{quotation}
If the above assumption is true, the centered inverse first moment, $E(\mathbf{X}|\mathbf{Y})-E(\mathbf{X})$ falls in a subspace of $\mathbf{R}^{p}$ spanned by $\Sigma \mathbb{B}$, where $\Sigma$ is covariance matrix of $\mathbf{X}$. A direct consequence of this is that the covariance matrix, $\mathbb{M}:= \mbox{Cov}\left\lbrace E(\mathbf{X}|\mathbf{Y})\right\rbrace$ is degenerate in any direction $\Sigma$-orthogonal to Span($\mathbb{B}$). Therefore, the eigenvectors corresponding to the $\mathit{d}$ nonzero eigenvalues of $\Sigma^{-1} \mathbf{M}$ span the subspace spanned by the columns of $\mathbb{B}$ and serves as the SIR estimates. To obtain an estimate for $E(\mathbf{X}|\mathbf{Y})$ when the response variable is quantitative, \citep{Li1991} suggested ordering the values of $\mathbf{Y}$, slicing it into non-overlapping ranges, and tuning into categories. The estimation procedure is provided in algorithm 1.

\setcounter{tblEqCounter}{\theequation} %at the start of the table, set the counter to equation numbering
\begin{table}[h]
%\caption{Basic Table}
\label{tab:siralg}
\centering
%\scriptsize
\begin{tabular}{llc p{5cm}}
\arrayrulecolor{blue} \toprule
\multicolumn{3}{l}{\textbf{Algorithm 1:} Sliced Inverse Regression (SIR)}   \\
\arrayrulecolor{blue} \midrule
 1. & Let $\mathbf{\widehat{\Sigma}},~\bar{\mathbf{X}}_{y}$ and $\bar{\mathbf{X}}$ be the sample versions of $\mathbf{\Sigma}, ~E(\mathbf{X}|\mathbf{Y})$ and $E(\mathbf{X})$, respectively.  &  \\ %labels are optional\\  
2. & Construct discretized versions of $\widetilde{\mathbf{Y}}$ of the response $\mathbf{Y}$ into $\mathbf{\mathit{h=1,2,\ldots, H}}$ \\
& approximately equal slices and obtain sample standardized predictor vector &\\
&~~~~~~~~~~~~~~~$\mathbf{Z}_{i} =\widehat{\mbox{var}}(\mathbf{X})^{\frac{1}{2}}(\mathbf{X}_{i}-\bar{\mathbf{X}}), i=1,\ldots, n$ &\numberTblEq{}\\ 
&where $\widehat{\mbox{var}}(X)$  is the usual estimate of the marginal covariance matrix of $\mathbf{X}$, & \\
& and $\bar{\mathbf{X}}$ is the sample mean of the predictor vector.&\\
3.& Construct the $p\times p~$ SIR kernel matrix& \\
&~~~~~~~~~~~~~~~$\widehat{\mathbb{M}}_{\mbox{SIR}}=\mathbf{\sum_{h=1}^{H} f_{h}}\bar{\mathbf{Z}}_{h}\bar{\mathbf{Z}}_{h}^{T}$  &\numberTblEq{eq3}\\
&where $f_{h}$ is the fraction of observations falling in slice $h$, and $\bar{\mathbf{Z}}_{h}$ is the &\\
& average of the sample standardized predictor vector. & \\ 
4.&Let $\mathbf{\hat{u}_{1},\ldots,\hat{u}_{p}}$ be eigenvectors of $\hat{\mathbf{M}}_{\mbox{SIR}}$ corresponding to its eigenvalues &\\
&$\mathbf{\hat{\lambda}_{1}\geq,\ldots,\geq \hat{\lambda}_{p}}$. Then the estimated coefficient vectors $\mathbf{\hat{\beta}}_{j}$ are again given as:&\\
& $~~~~~~~~~~~~~\hat{\mathbf{\beta}}_{j} = \widehat{\mbox{var}}(\mathbf{X})^{\mathbf{\frac{1}{2}}}\mathbf{\hat{u}_{j}},~~~~~~~~      j=1,\ldots,p.$&\numberTblEq{}\\
& The corresponding predictors $\widehat{\mathbf{\beta}}_{j}^{T} \mathbf{X}, j=1,\ldots,p$, are called the SIR predictors.&\\ \arrayrulecolor{blue}
\bottomrule
\end{tabular}
\end{table}
\setcounter{equation}{\thetblEqCounter} %at the end of the table, set the equation numbering to the counter
\cite{Li1991} noted that SIR cannot find more than $\mathbf{\mathit{H-1}}$ sufficient predictors. If $\mathbf{\mathit{H< d+1}}$ then the set of sufficient predictors for $\mathbf{\widetilde{Y}}$ on $\mathbf{X}$ will necessarily exclude some of the sufficient predictors for the regression of $\mathbf{Y}$ on $\mathbf{X}$. Good results are often obtained by choosing $\mathbf{\mathit{H}}$ to be somewhat larger than $\mathbf{\mathit{d}+1}$. If possible, trying a few different values of $\mathbf{\mathit{H}}$ is necessary. Choosing $\mathbf{\mathit{H}}$ substantially larger than $\mathit{d}$ should be avoided since it can lead to $\mathbf{Y} =\mathbf{\widetilde{Y}}$ (see \citeauthor{Cook2003}, \citeyear[sec. 4.2]{Cook2003}; \citeauthor{Cook1998}, \citeyear[cap 11]{Cook1998}; and \citeauthor{Li1991}, \citeyear{Li1991}). Over the last two decades, the classical SIR has been studied extensively and it remains the most popular sufficient dimension reduction method despite the plethora of sufficient dimension reduction method proposals \cite{Hilafu2017}. 
\subsection{Sliced Average Variance Estimation}
\label{subsec:SAVE}
The pioneering SIR is a first-order method based on linear conditional mean assumption that $E(\mathbf{X}|\mathbb{B}^{T} \mathbf{X})$ is a linear function of $\mathbb{B}^{T} \mathbf{X}$. Extensive studies have shown that it cannot recover any vector in the central subspace $\mathbf{S}_{(\mathbf{Y}|\mathbf{X})}$, if the regression function is symmetric about 0.
\setcounter{tblEqCounter}{\theequation} %at the start of the table, set the counter to equation numbering
\begin{table}[ht!]
%\caption{Basic Table}
\label{tab:savealg}
\centering
%\scriptsize
\begin{tabular}{llc p{5cm}}
\arrayrulecolor{blue} \toprule
\multicolumn{3}{l}{\textbf{Algorithm 2:} Sliced Average Variance Estimation (SAVE)}   \\ \arrayrulecolor{blue}
\midrule
 1. & 	Construct discretized version $\widetilde{\mathbf{Y}}$ of the bivariate response, $\mathbf{Y}$ by slicing into & \\
 &$\mathbf{\mathit{h=1,2,\ldots}},\mathbf{\mathit{H}}$ approximately equal slices and obtain sample standardized&\\& predictor vector &  \\ %labels are optional
 &~~~~~~~~~~~~~~$\mathbf{Z}_{i}=\widehat{\mbox{var}}(\mathbf{X})^{\mathbf{-\frac{1}{2}}} (\mathbf{X}_{i}-\bar{\mathbf{X}}),~~i=1,\ldots,n$& \numberTblEq{}\\  
 & where $\widehat{\mbox{var}}(\mathbf{X})$ is the usual estimate of the marginal covariance matrix of $\mathbf{X}$,&\\ 
 &and $\bar{\mathbf{X}}$ is the sample mean of the predictor vector.&\\
2. & 	Construct the $p\times p$ kernel matrix  &\\
& ~~~~~~~~~~~~~~~$\widehat{\mathbb{M}}_{\mbox{SAVE}}=\mathbf{\sum_{h=1}^{H} f_{h}(1-\hat{\sum}_{h})^{2}}$&\numberTblEq{}\\ 
&where $\mathbf{\hat{\sum}_{h}}$ denote the estimated covariance matrix for the vector of standard-&\\
& ized predictors within slice h. & \\
3.&Let $\mathbf{\hat{u}_{1},\ldots,\hat{u}_{p}}$ be eigenvectors of $\hat{\mathbf{M}
}_{\mbox{SAVE}}$ corresponding to its eigenvalues &\\
&$\mathbf{\hat{\lambda}_{1} \geq,\ldots,\geq \hat{\lambda}_{p}}$. Then the estimated coefficient vectors $\widehat{\mathbf{\beta}}_{j}$ are again given by & \\
& ~~~~~~~~~~~~~~~~$\widehat{\mathbf{\beta}}_{j}=\widehat{\mbox{var}}(\mathbf{X})^{-\mathbf{\frac{1}{2}}}\hat{u}_{j},~~~  j=1,\ldots,p.$&\numberTblEq{}\\
& The corresponding predictors  $\widehat{\mathbf{\beta}}_{j}^{T}\mathbf{X},~~j=1,\ldots,p$, are called the \textit{SAVE} predictors.&\\ \arrayrulecolor{blue}
\bottomrule
\end{tabular}
\end{table}
\setcounter{equation}{\thetblEqCounter} \\
To remedy the situation, methods based on the second-order conditional moments (second-order methods), such as $\mbox{Var}(\mathbf{X}|\mathbf{Y})$  and $E(\mathbf{XX}^{T} |\mathbf{Y})$, have been developed. Sliced Average Variance Estimation (SAVE; \citeauthor{CookWeisberg1991}, \citeyear{CookWeisberg1991}) method is the first of such methods. In addition to the linearity condition in SIR, SAVE is also based on constant conditional variance assumption that $\mbox{Var}(\mathbf{X}|\mathbb{B}^{T} \mathbf{X})$ is a nonrandom matrix. Let $(\mathbf{X}_{1},\mathbf{Y}_{1}),\ldots ,(\mathbf{X}_{n},\mathbf{Y}_{n})$ be an independent sample of $(\mathbf{X},\mathbf{Y})$, we provide estimation procedure of SAVE in algorithm 2. \\
SAVE was further developed by \cite{CookLee1999} and  \cite{CookCritchley2000}. An account of basic methodology was given by  \cite{Cook2003}. Like SIR, a plot of $\mathbf{Y}$ versus the first two SAVE predictors and a marked plot of $\mathbf{\widehat{Y}}$ versus the first three SAVE predictors are usually informative in practice \citep{Cook2003}. 
\section{Maximum Entropy Covariance Selection (MECS) Method}
\label{sec:mecs}
MECS was proposed by \cite{Thomaz2002,Thomaz2004} to circumvent limited-sample-size problem for Bayesian classifiers in biometric recognition. MECS method was built on the \textquotedblleft loss of covariance information\textquotedblright paradigm and the maximum entropy (ME) principle.
\subsection{Loss of Covariance Information}
\label{sec:lci}
We first describe the \textquotedblleft loss of covariance information\textquotedblright paradigm in limited-sample-size problem.\\
\textbf{\textit{Proposition 3.1:}}  Let $\mathbf{S}_{i}$ and $\mathbf{S}_{p}$ denote the unbiased maximum-likelihood estimators of the true samples group covariance matrices and the pooled sample group covariance matrix defined as
\begin{equation}\label{eq:8}
\mathbf{S}_{i} = \frac{1}{(n_{i}-1)}\sum_{j=1}^{n_{i}}(x_{i,j}-\bar{x}_{i})(x_{i,j}-\bar{x}_{i})^{T}
\end{equation}
and 
\begin{equation}\label{eq:9}
\mathbf{S}_{p} = \frac{(n_{1}-1)\mathbf{S}_{1}+(n_{2}-1)\mathbf{S}_{2}+\cdots +(n_{g}-1)\mathbf{S}_{g}}{N-g}
\end{equation}
where $x_{i,j}$ is the pattern $\mathit{j}$ from class $i=1,\ldots,g,~  n_{i}$  is the number of training patterns from class $i,~ g$ is the number of classes and $N=n_{1}+n_{2}+\cdots+n_{g}$. The \textquotedblleft loss of covariance information\textquotedblright  can be described by mixture covariance matrix $\mathbf{S}_{i}^{\mbox{mix}}$ given by the linear combination,
\begin{equation}\label{eq:10}
\mathbf{S}_{i}^{\mbox{mix}} = a\mathbf{S}_{i}+b\mathbf{S}_{p}
\end{equation}
where the mixing parameters $\mathit{a}$  and $\mathit{b}$ are positive constants, the proof is straightforward in terms of the sample group covariance matrix spectra decomposition formula: 
\begin{equation}\label{eq:11}
\mathbf{S}_{i} = \mathbf{\Psi}_{i}\Lambda_{i}\mathbf{\Psi}_{i}^{T} =\sum_{k=1}^{p} \mathbf{\lambda}_{ik}\psi_{ik}\mathbf{\psi}_{ik}^{T},
\end{equation}
and its inverse covariance matrix defined as: 
\begin{equation}\label{eq:12}
\mathbf{S}_{i}^{-1} =\sum_{k=1}^{p}\frac{\mathbf{\psi}_{ik}\mathbf{\psi}_{ik}^{T}}{\mathbf{\lambda}_{ik}},
\end{equation}
where $\mathbf{\lambda}_{ik}$ is the $kth$ eigenvalue of $\mathbf{S}_{i}$  , $\mathbf{\psi}_{ik}$  is the corresponding eigenvector, $\mathbf{\Psi}_{i}$ and $\mathbf{\Lambda}_{i}$  are the corresponding eigenvector and eigenvalue matrices of $\mathbf{S}_{i}$, respectively. 
Without loss of generality, the covariance spectra decomposition formula in (\ref{eq:11}), can be defined as:
\begin{equation}\label{eq:13}
\left(\mathbf{\Psi}_{i}^{\mbox{mix}}\right)^{T} \mathbf{S}_{i}^{\mbox{mix}}\mathbf{\Psi}_{i}^{\mbox{mix}}=\Lambda_{i}^{mix} =\left[ \begin{array}{cccc}
\mathbf{\Psi}_{1}^{\mbox{mix}} &  &  & 0 \\ 
  & \mathbf{\Psi}_{2}^{\mbox{mix}} &   &  \\ 
  &   &  \ddots &  \\ 
0 &   &   & \mathbf{\Psi}_{p}^{\mbox{mix}}
\end{array}\right],
\end{equation}
where $\mathbf{\lambda}_{1}^{\mbox{mix}},\mathbf{\lambda}_{2}^{\mbox{mix}},\ldots, \mathbf{\lambda}_{p}^{\mbox{mix}}$  are the eigenvalues of the sample group mixture covariance matrix $\mathbf{S}_{i}^{\mbox{mix}}$  and $\mathit{p}$ is the dimension of the measurement space considered. Plugging (\ref{eq:10}) and (\ref{eq:13}) together, a direct implication of (\ref{eq:10}) is that 
\begin{subequations}\label{eq:14}
\begin{align}
\label{eq:14a}
\mathbf{\Lambda}_{i}^{\mbox{mix}} &= \mbox{diag}\left[\mathbf{\lambda}_{1}^{\mbox{mix}}, \mathbf{\lambda}_{2}^{\mbox{mix}}, \ldots, \mathbf{\lambda}_{p}^{\mbox{mix}}\right] \nonumber\\
&= \left(\mathbf{\Psi}_{i}^{\mbox{mix}}\right)^{T}\left[a\mathbf{S}_{i}+b\mathbf{S}_{p}\right]\mathbf{\Psi}_{i}^{\mbox{mix}}\nonumber \\
&= a\left(\mathbf{\Psi}_{i}^{\mbox{mix}}\right)^{T}\mathbf{S}_{i}\mathbf{\Psi}_{i}^{\mbox{mix}}+b\left(\mathbf{\Psi}_{i}^{\mbox{mix}}\right)^{T}\mathbf{S}_{p}\mathbf{\Psi}_{i}^{\mbox{mix}}\nonumber\\
&= a\mathbf{Z}^{i}+b\mathbf{Z}^{p} 
\end{align}
The matrix $\mathbf{\Psi}_{i}^{\mbox{mix}}$ is the eigenvectors matrix of the linear combination of $\mathbf{S}_{i}$  and $\mathbf{S}_{p}$  . The off-diagonal elements of $\mathbf{Z}^{i}$ and $\mathbf{Z}^{p}$ necessarily cancel each other in order to generate the diagonal matrix of eigenvalues $\mathbf{\Lambda}_{i}^{\mbox{mix}}$ \citep{Thomaz2004}.

\begin{corollary}\label{cor:1}
Suppose that $\mathbf{X}_{i},~i=1,\ldots,g$ are normally distributed and $\mathbf{S}_{i}$ and $\mathbf{S}_{p}$ denote the unbiased maximum-likelihood estimators of true sample group covariance and pooled sample covariance matrices. Then the eigenvalue matrix in (\ref{eq:14a}) is equivalent to linear combination of variances of $\mathbf{S}_{i}$ and $\mathbf{S}_{p}$ spanned by the $\mathbf{S}_{i}^{\mbox{mix}}$ eigenvectors matrix $\mathbf{\Psi}_{i}^{\mbox{mix}}$.
\end{corollary}

From corollary \ref{cor:1}, equation (\ref{eq:14a}) can be extended to 
\begin{align}
\label{eq:14b}
\mathbf{\Lambda}_{i}^{\mbox{mix}} &= a\mathbf{Z}^{i}+b\mathbf{Z}^{p}\nonumber\\
&= \mbox{diag}\left[ a\varphi_{1}^{i}, a\varphi_{2}^{i},\ldots, a\varphi_{p}^{i} \right]+\mbox{diag}\left[ b\varphi_{1}^{p},b\varphi_{2}^{p}, \ldots,b\varphi_{p}^{p} \right] \nonumber\\
&= \mbox{diag}\left[ a\varphi_{1}^{i}+b\varphi_{1}^{p},a\varphi_{2}^{i}+b\varphi_{2}^{p},\ldots, a\varphi_{p}^{i}+b\varphi_{p}^{p} \right]
\end{align}
\end{subequations}
where $\mathbf{\varphi}_{1}^{i},\mathbf{\varphi}_{2}^{i},\ldots,\mathbf{\varphi}_{p}^{i}$ and $\mathbf{\varphi}_{1}^{p},\mathbf{\varphi}_{2}^{p},\ldots,\mathbf{\varphi}_{p}^{p}$ are, respectively, the variances of the sample and pooled covariance matrices spanned by the $\mathbf{S}_{i}^{\mbox{mix}}$ eigenvectors of $\mathbf{\Psi}_{i}^{\mbox{mix}}$.\\
Then it follows from (\ref{eq:12}) that the inverse of $\mathbf{S}_{i}^{\mbox{mix}}$ becomes
\begin{equation}\label{eq:15}
\left(\mathbf{S}_{i}^{-1}\right)^{-1} = \sum_{k=1}^{p}\frac{\mathbf{\psi}_{ik}^{\mbox{mix}}\left(\mathbf{\psi}_{ik}^{\mbox{mix}}\right)^{T}}{a\mathbf{\psi}_{k}^{i}+b\mathbf{\psi}_{k}^{p}}
\end{equation}
The inverse of $\mathbf{S}_{i}^{\mbox{mix}}$ described in (\ref{eq:15}) considers the dispersions of sample group covariance matrices spanned by all the $\mathbf{S}_{i}^{\mbox{mix}}$ eigenvectors. However, when the class sample size $n_{i}$ are undersized compared to p, the corresponding lower dispersion values are often estimated to be zero or approximately so, implying that these values are not reliable. Therefore, a linear combination of $\mathbf{S}_{i}$ and $\mathbf{S}_{p}$ that uses the same parameters $\mathit{a}$ and $\mathit{b}$ as defined in (\ref{eq:10}) for the whole feature space fritters away some pooled covariance information. As a direct consequence of \textquotedblleft loss of covariance information\textquotedblright, ~a linear combination of $\mathbf{S}_{i}$ and $\mathbf{S}_{p}$ that shrinks or expands both matrices equally all over the features space simply ignores this evidence (see \citeauthor{Thomaz2004}, \citeyear{Thomaz2004}). All covariance estimators lack the ability to address \textquotedblleft loss of covariance information\textquotedblright in their estimation procedure, except those based on maximum uncertainty (entropy) principle.
\subsection{Maximum Entropy Principle}
\label{sec:mepriniple}
The principle of maximum entropy (ME) states that 
\begin{quote}
 \textquotedblleft The probability distribution which best represents the current state of knowledge is the one with largest entropy\textquotedblright
\end{quote}  
The implication of the ME principle is that: when we make inferences based on incomplete information, we should draw them from that probability distribution that has the maximum entropy permitted by the information we do have ~\citep{Jaynes1982}. \\
In the problem of estimating covariance matrices for Gaussian classifiers, it is known that different covariance estimators should be optimal depending not only on the true covariance statistics of each class, but also on the number of training patterns, the dimension of the feature space, and even the elliptical symmetry associated with the Gaussian distribution ~\citep{James1985, Friedman1989}. Since entropy is the average rate at which information is produced from a stochastic source of data and covariance optimization can be viewed as a problem of estimating parameters of Gaussian probability distributions under uncertainty, the ME criterion that minimizes \textquotedblleft loss of covariance information\textquotedblright by maximizing the uncertainty under an incomplete information context should be a promising solution.\\
Let an p-dimensional sample $\mathbf{X}_{i}$ of class probability $\mathbf{\pi}_{i}$ be normally distributed with true mean $\mathbf{\mu}_{i}$ and true covariance matrix $\mathbf{\Sigma}_{i},$~ i.e.$\mathbf{X}_{i}\sim \mathbf{N}_{p} \left(\mathbf{\mu}_{i},\mathbf{\Sigma}_{i}\right)$. The entropy $\mathbf{h}\left(\mathbf{X}_{i}\right)$ of such multivariate distribution is defined as the expected value of the natural logarithm of the probability density function of $\mathbf{X}_{i}$, which can be written as~\citep{Fukunaga1990}.
\begin{eqnarray}\label{eq:16}
h(X_{i}) &=& -E\left\lbrace \ln \left[p(x|\pi_{i})\right] \right\rbrace \nonumber\\
&=&  -E\left\lbrace\ln \left[\frac{1}{(2\pi)^{\frac{p}{2}} |\Sigma_{i}|^{\frac{1}{2}}}\exp\left[-\frac{1}{2}(x-\mu_{i})^{T}\Sigma_{i}^{-1} (x-\mu_{i}) \right] \right]\right\rbrace \nonumber\\
&=& -E\left\lbrace -\frac{p}{2}\ln (2\pi) -\frac{1}{2} \ln \vert \Sigma_{i}\vert -\frac{1}{2}(x-\mu_{i})^{T}\Sigma_{i}^{-1}(x-\mu_{i})\right\rbrace \nonumber\\
&=& -E\left\lbrace -\frac{p}{2}\ln (2\pi) \rbrace-E\lbrace -\frac{1}{2}\ln \vert \Sigma_{i}\vert\rbrace -E\lbrace -\frac{1}{2}(x-\mu_{i})^{T}\Sigma_{i}^{-1}(x-\mu_{i}) \right\rbrace \nonumber\\
&=& \frac{p}{2}\ln 2\pi +\frac{1}{2}\ln \vert\Sigma_{i} \vert +\frac{p}{2}. 
\end{eqnarray}
By dropping the constant terms $(p/2)ln 2 \mathbf{\pi}$ and $p/2$, the entropy $\mathbf{h}(\mathbf{X}_{i})$ is simply a function of the determinant $\mathbf{\Sigma}_{i}$, which is invariant under any orthonormal transformation \citep{Fukunaga1990}. 
\subsection{The MECS Algorithm}
\label{sec:mecsalg}
The MECS method considers the issue of convex combination of the sample group covariance matrices and the pooled covariance matrix to address \textquotedblleft loss of covariance information\textquotedblright in limited sample-size problems using maximum entropy principle. By assuming that all classes have similar covariance shapes, it is reasonable to expect that the dominant eigenvectors (i.e. the eigenvectors with largest eigenvalues) of this unbiased mixture would be mostly oriented by the eigenvectors of the covariance matrix with largest eigenvalues~ \citep{Thomaz2004}.
Thus, $\mathbf{\Psi}_{i}$ consists of $\mathit{p}$ eigenvectors of $\mathbf{\Sigma}_{i}$, we have 
\begin{equation}\label{eq:17}
ln\left| \mathbf{\Psi}_{i}^{T} \mathbf{\Sigma}_{i} \mathbf{\Psi}_{i} \right| =ln \left| \mathbf{\Lambda}_{i} \right|= \sum_{k=1}^{p} ln \lambda_{k}.                                                                    
\end{equation}
In order to maximize the entropy (\ref{eq:16}) or equivalently (\ref{eq:17}), the covariance estimation of $\mathbf{\sigma}_{i}$ that gives the largest eigenvalues must be selected \citep{Thomaz2004}. By convex combination of $\mathbf{S}_{i}$ and $\mathbf{S}_{p}$ matrices, (\ref{eq:17}) can be rewritten as:
\begin{equation}\label{eq:18}
ln  \left|\left(\mathbf{\Psi}_{i}^{\mbox{mix}}\right)^{T} \left(a\mathbf{S}_{i}+b\mathbf{S}_{p}\right) \mathbf{\Psi}_{i}^{\mbox{mix}} \right| = \sum_{k=1}^{p} ln \left(a\mathbf{\varphi}_{k}^{i}+b\mathbf{\varphi}_{k}^{p}\right)
\end{equation}
where $\mathbf{\varphi}_{1}^{i},\mathbf{\varphi}_{2}^{i},\ldots,\mathbf{\varphi}_{p}^{i}$ and $\mathbf{\varphi}_{1}^{p},\mathbf{\varphi}_{2}^{p},\ldots,\mathbf{\varphi}_{p}^{p}$ are the variances of the sample and pooled covariance matrices spanned by $\mathbf{\Psi}_{i}^{\mbox{mix}}$, and the parameters a and b are nonnegative and sum to 1.\\
Moreover, because natural logarithm is a monotonic increasing function,\cite{Thomaz2004} stated that the problem remains unchanged if instead of maximizing (\ref{eq:18}), we maximize
\begin{equation}\label{eq:19}
\sum_{k=1}^{p} \left(a\mathbf{\varphi}_{k=1}^{i}+b\mathbf{\varphi}_{k}^{p}\right).
\end{equation}
However, $a\mathbf{\varphi}_{k}^{i}+b\mathbf{\varphi}_{k}^{p}$ is a convex combination of two real numbers and the following inequality is valid \citep{HornJonson1985}:
\begin{equation}\label{eq:20}
a \mathbf{\varphi}_{k}^{i} + b \mathbf{\varphi}_{k}^{p} \leq\max\mathbf{\varphi}_{k}^{i}+\mathbf{\varphi}_{k}^{p})
\end{equation}
for any $1 \leq k \leq p$ and convex parameters $\mathit{a}$ and $\mathit{b}$. The consequence of equation (\ref{eq:20}) is that the maximum of $a\mathbf{\varphi}_{k}^{i}+b\mathbf{\varphi}_{k}^{p}$ depends on $\mathit{k}$ and is attained at the extreme values of the convex parameters, that is, either $\mathit{a}=1$ and $\mathit{b}=0$ or $\mathit{a}=0$ and $\mathit{b}=1$.
Therefore, the MECS estimator $\mathbf{S}_{i}^{\mbox{mecs}}$ can be calculated by the procedure outlined in algorithm 3.
\setcounter{tblEqCounter}{\theequation} %at the start of the table, set the counter to equation numbering
\begin{table}[ht!]
%\caption{Basic Table}
\label{tab:mecsalg}
\centering
%\scriptsize
\begin{tabular}{llc p{5cm}}
\arrayrulecolor{blue} \toprule
\multicolumn{3}{l}{\textbf{Algorithm 3:} Maximum Entropy Covariance Selection (MECS) METHOD}   \\ \arrayrulecolor{blue}
\midrule
 1. & 	Find the eigenvectors $\mathbf{\Psi}_{i}^{\mbox{me}}$ of the covariance given by $\mathbf{S}_{i} + \mathbf{S}_{p}$.& \\
2. &Calculate the variance contribution of both $\mathbf{S}_{i}$ and $\mathbf{S}_{p}$ on the $\mathbf{\Psi}_{i}^{me}$ basis, i.e.,&\\
&~~~~~~~~~~~~~~~$\mbox{diag}\left(\mathbf{Z}^{i} \right) = \mbox{diag}\left[ \left(\mathbf{\Psi}_{i}^{\mbox{me}}\right)^{T}\mathbf{S}_{i}\mathbf{\Psi}_{i}^{\mbox{me}}\right]=\left[\mathbf{\varphi}_{1}^{i}, \mathbf{\varphi}_{2}^{i},\ldots, \mathbf{\varphi}_{p}^{i}\right]$&\\
& ~~~~~~~~~~~~~~$\mbox{diag}\left(\mathbf{Z}^{p} \right) = \mbox{diag}\left[ \left(\mathbf{\Psi}_{p}^{\mbox{me}}\right)^{T}\mathbf{S}_{p}\mathbf{\Psi}_{p}^{\mbox{me}}\right]=\left[\mathbf{\varphi}_{1}^{p}, \mathbf{\varphi}_{2}^{p},\ldots, \mathbf{\varphi}_{p}^{p}\right]$&\numberTblEq{}\\
3.&Form a new variance matrix based on the largest values, that is,&\\
& ~~~~~~~~~~~~~~$\mathbf{Z}_{i}^{\mbox{me}}=\mbox{diag} \left[\max(\mathbf{\varphi}_{1}^{i},\mathbf{\varphi}_{1}^{p} ),\ldots, \max(\mathbf{\varphi}_{p}^{i},\mathbf{\varphi}_{p}^{p})\right]$&\numberTblEq{}\\
4. &Form the MECS estimator&\\
& ~~~~~~~~~~~~~~$\mathbf{S}_{i}^{\mbox{mecs}} = \mathbf{\Psi}_{i}^{\mbox{me}}\mathbf{Z}_{i}^{\mbox{me}}\left(\mathbf{\Psi}_{i}^{\mbox{me}}\right)^{T}$&\numberTblEq{}\\ \arrayrulecolor{blue}
\bottomrule
\end{tabular}
\end{table}
\setcounter{equation}{\thetblEqCounter} %at the end of the table, set the equation numbering to the counter
MECS is a direct procedure that not only deals with the singularity and instability of $\mathbf{S}_{i}$ but also with the loss of information when similar covariance matrices are linearly combined. Because it does not require time-consuming covariance optimization procedure, its computational cost is much less severe than many popular covariance estimator designed for solving limited-sample-size problems in quadratic discriminant classifiers \citep{Thomaz2004}. 
\section{Proposed Maximum Entropy Covariance (MEC) Estimator}
\label{sec:mec}
A new covariance estimator called Maximum Entropy Covariance (MEC) estimator is proposed to address the present potentially daunting limitations of the existing MECS estimator. (1) in sufficient dimension reduction regression applications involving undersized samples, high-dimensional predictors and quantitative response variable $\mathbf{Y}$ where sample groups are not known a priori in the data but rather by slicing or discretizing the range of Y into approximately equal and $\mathit{H}$ non-overlapping groups, and (2) in classification and regression analyses with data sets involving undersized samples and ultrahigh-dimensional predictors where the existing MECS method under-utilizes the Maximum Entropy (ME) principle and often breaks down.\\
The proposed MEC method addressed these limitations of the MECS method on accounts of the reliable consequence of the ME principle that among all sample group covariance matrices, $\mathbf{S}_{i},~i=1,2,\ldots,g$,  one such sample group covariance matrix denoted $\mathbf{S}_{\mbox{me}}$ maximizes the eigenvalue and hence contains the most reliable information, without loss of information and generality. Hence, MEC deals with loss of covariance information in (\ref{eq:10}) by redefining the covariance mixture $\mathbf{S}_{i}^{\mbox{mix}}$ as:
\begin{equation}\label{eq:24}
\mathbf{S}^{\mbox{mix}}=a\mathbf{S}_{\mbox{me}}+b\mathbf{S}_{p}. 
\end{equation} 
Considering convex combination of $\mathbf{S}_{\mbox{me}}$ and $\mathbf{S}_{p}$ matrices in (\ref{eq:24}), (\ref{eq:17}) becomes:
\begin{equation}\label{eq:25}
ln \left|\left(\mathbf{\Psi}^{\mbox{mix}} \right)(\mathit{a}\mathbf{S}_{\mbox{me}}+\mathit{b}\mathbf{S}_{\mathit{p}})\mathbf{\Psi}^{\mbox{mix}}\right| = \sum_{k=1}^{\mathit{p}}ln\left(\mathit{a}\mathbf{\varphi}_{k}^{\mbox{me}}+\mathit{b}\mathbf{\varphi}_{k}^{\mathit{p}} \right),
\end{equation}
where $\mathbf{\varphi}_{1}^{\mbox{me}}, \mathbf{\varphi}_{2}^{\mbox{me}},\ldots,\mathbf{\varphi}_{p}^{\mbox{me}}$ and $\mathbf{\varphi}_{1}^{p},\mathbf{\varphi}_{2}^{p},\ldots,\mathbf{\varphi}_{p}^{p}$ are the variances of the maximum entropic sample group and pooled sample group covariance matrices spanned by $\mathbf{\Psi}^{\mbox{mix}}$, and the parameters $\mathit{a}$ and $\mathit{b}$ are nonnegative and summed to 1. 
\setcounter{tblEqCounter}{\theequation} 
\begin{table}[ht!]
%\caption{Basic Table}
%\label{tab:mecalg}
\centering 
%\scriptsize
\begin{tabular}{llc p{5cm}}
\arrayrulecolor{blue} \toprule
\multicolumn{3}{l}{\textbf{Algorithm 4:} Maximum Entropy Covariance (MEC) Estimator}   \\ \arrayrulecolor{blue}
\midrule
 1. & 	Find covariance $\mathbf{S}_{i}$ for each sample group i (or slicing category $\mathit{h}$) and the pooled &\\
 &sample covariance matrix $\mathbf{S}_{p}$.& \\
2. &Find the maximum entropic sample group covariance estimate $\mathbf{S}_{\mbox{me}}$.&\\
3.&Find the eigenvectors $\mathbf{\Psi}_{i}^{\mbox{me}}$  of the convex covariance mixture given by $\mathbf{S}_{\mbox{me}}+\mathbf{S}_{p}$.&\\
4. & Calculate the variance contribution of both $\mathbf{S}_{\mbox{me}}$ and $\mathbf{S}_{p}$ on the $\mathbf{\Psi}^{\mbox{me}}$ basis, i.e.,&\\
& $\mbox{diag}(\mathbf{Z}^{\mbox{me}})=\mbox{diag}\left[(\mathbf{\Psi}^{\mbox{me}})^{T} \mathbf{S}_{\mbox{me}} \mathbf{\Psi}^{\mbox{me}} \right]=\left[\mathbf{\varphi}_{1}^{\mbox{me}},\mathbf{\varphi}_{2}^{\mbox{me}},\ldots,\mathbf{\varphi}_{p}^{\mbox{me}} \right]$&\\
& ~~~~~~~~~~~~~$\mbox{diag}(\mathbf{Z}^{p})=\mbox{diag}\left[(\mathbf{\Psi}^{\mbox{me}} )^{T} \mathbf{S}_{p} \mathbf{\Psi}^{\mbox{me}}\right]=\left[\mathbf{\varphi}_{1}^{p},\mathbf{\varphi}_{2}^{p},\ldots,\mathbf{\varphi}_{p}^{p} \right]$.&\numberTblEq{}\\
5. & Form a new variance matrix based on the largest values, that is,&\\
& ~~~~~~~~~~~~~$\mathbf{Z}_{i}^{\mbox{me}}=\mbox{diag}\left[\max  \left(\mbox{mean}(\mathbf{\varphi}_{1}^{\mbox{me}},\mathbf{\varphi}_{1}^{p})\right),\ldots,\max  (\mbox{mean}(\mathbf{\varphi}_{p}^{\mbox{me}},\mathbf{\varphi}_{p}^{p}))\right]$&\numberTblEq{}\\
6. &	Form the MEC estimator, $\mathbf{S}^{\mbox{mec}}$ as:&\\
& ~~~~~~~~~~~~~$\mathbf{S}^{\mbox{mec}}=\mathbf{\Psi}^{\mbox{me}} \mathbf{Z}^{\mbox{me}} (\mathbf{\Psi}^{\mbox{me}})^{T}$.&\numberTblEq{}\\ \arrayrulecolor{blue}
\bottomrule
\end{tabular}
\end{table}
\setcounter{equation}{\thetblEqCounter} %at the end of the table, set the equation numbering to the counter
\noindent
Since natural logarithm is a monotonic increasing function, the problem remains unchanged if instead of maximizing (\ref{eq:18}),
\begin{equation}\label{eq:26}
\sum_{k=1}^{p} \left(a\mathbf{\varphi}_{k}^{\mbox{me}}+b\mathbf{\varphi}_{k}^{p}\right)
\end{equation}
is maximized. However, $a\mathbf{\varphi}_{k}^{\mbox{me}}+b\mathbf{\varphi}_{k}^{p}$ is a convex combination of two real numbers and the following inequality:
\begin{equation}\label{eq:27}
a\mathbf{\varphi}_{k}^{\mbox{me}}+b\mathbf{\varphi}_{k}^{p} \leq \max \left[ mean(\mathbf{\varphi}_{k}^{\mbox{me}}+\mathbf{\varphi}_{k}^{p})\right]
\end{equation}
still holds for any $1\leq k \leq p$ and convex parameters $\mathit{a}$ and $\mathit{b}$. The consequence of equation (\ref{eq:27}) is that the maximum of the average of the mixture in (\ref{eq:24}), $\mbox{mean}(a\mathbf{\varphi}_{k}^{\mbox{me}}+b\mathbf{\varphi}_{k}^{p} )$, depends on $\mathit{k}$ and is attained at the extreme values of the convex parameters, that is, either $\mathit{a}=1$ and $\mathit{b}=0$ or $\mathit{a}=0$ and $\mathit{b}=1$. Therefore, the MEC estimator, $\mathbf{S}^{\mbox{mec}}$ can be calculated by the procedure outlined in algorithm 4.

\section{Application and Results}
\label{sec:application}
In order to investigate the performance of MEC estimator in sufficient dimension reduction applications with SIR and SAVE, two example applications including binary discrimination of patients according to Leukemia cancer tumor statuses and a regression application including prediction of electricity customers' consumption profile were considered. The data sets utilized for the two applications are described in section \ref{sec:data}. 
Six standard statistical-based classifiers including Logistic Regression (LR)classifier, Linear Discriminant Analysis (LDA) classifier, Quadratic Discriminant Analysis (LDA) classifier, Naïve Bayes (NB) classifier, k-Nearest Neibour (k-NN) classifier, and Classification Trees (CTree), respectively, were trained and tested for classification performances with only first SDR direction of the SIR and SAVE. The data sets utilized for the two applications are described below.
\subsection{Data Description}
\label{sec:data} 
\begin{figure}[ht!]
\centering
\includegraphics[height = 16cm]{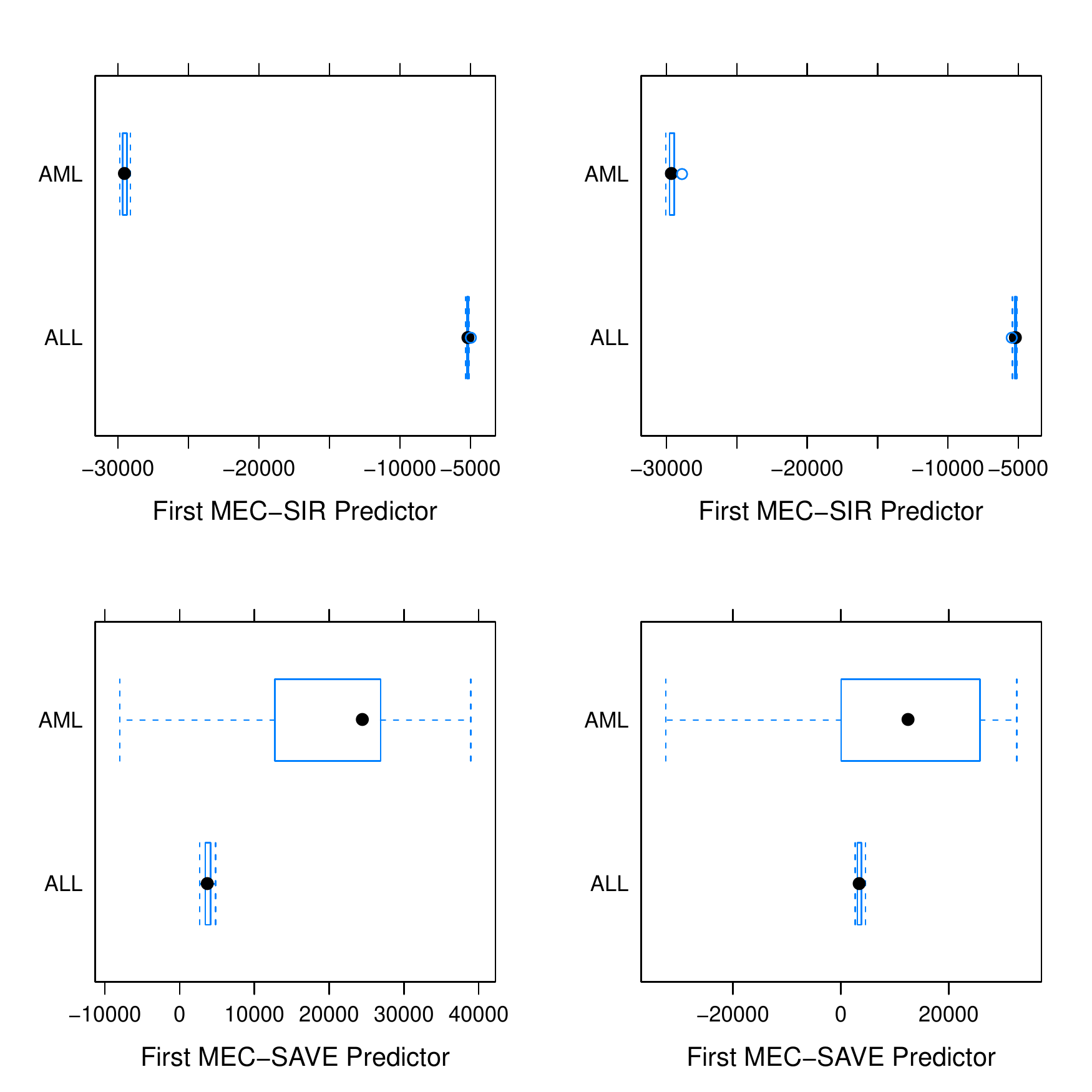}
\caption{SIR and SAVE estimates for the Leukemia data. Top panel is for the SIR and bottom panel is for SAVE}
\label{fig:1}
\end{figure}

\begin{figure}[ht!]
\centering
\includegraphics[height =14.5cm, width = 16cm]{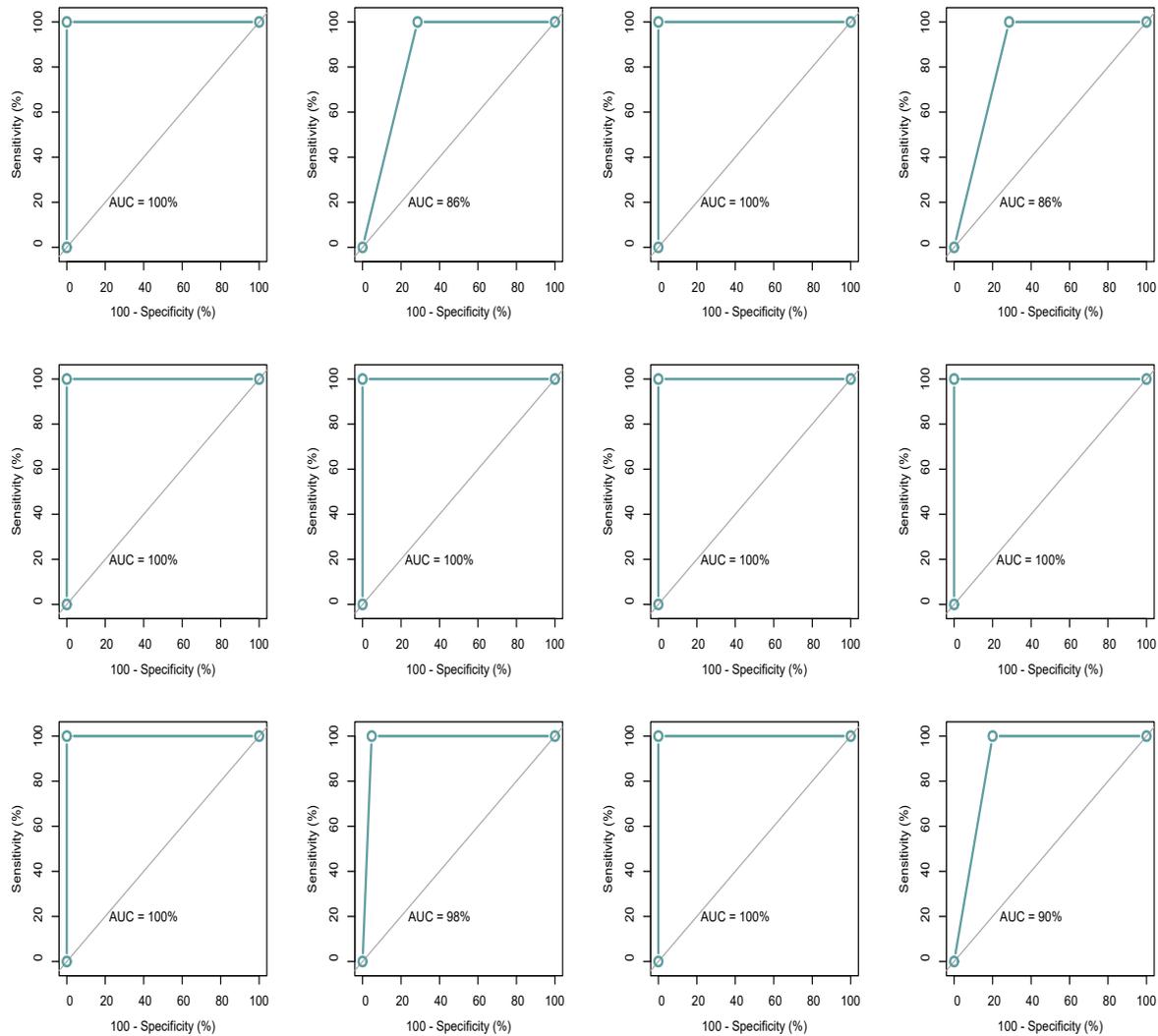}
\caption{Leukemia status classification performances with first MEC based SIR and SAVE predictors. Columns 1 and 3 are based on first SIR predictor while columns 2 and 4 are based on first SAVE predictor. The plots in row one represent Logistic Regression and Linear Discriminant Analysis performances with first SIR and SAVE predictors, respectively. Plots in middle row represent performances of QDA and 1-NN. Plots in last row represent performances of Naïve Bayes and Classification trees, respectively.}
\label{fig:2}
\end{figure}

\textbf{Leukemia cancer data (Leuk):} this data set is introduced by \cite{Golub1999}. It contains the expression levels of 7129 genes for 47 acute \textit{lymphoblastic leukemia (ALL)} patients and 25 \textit{acute myeloid leukemia (AML)} patients. \textit{ALL} arises from two different types of lymphocytes (T-cell and B-cell), the data set is usually considered in terms two classes or three classes: \textit{AML}, \textit{ALL-T}, and \textit{ALL-B}. The data were considered in terms two classes in our application. The test set comprises first 34 samples while the remaining 38 samples make up the training samples. It is a popular benchmark data set preloaded with any version of the R software for statistical computing and graphics \citep{R2019} under the \textit{datamicroarray} package \citep{Ramey2016}.\\
\noindent
\textbf{Smart Meter Data (SMD):} This data set comprised the raw $56\times 2975$ data matrix of electricity consumption records automatically generated by the smart meters. Each of the 2975 quarter-hourly electricity load records constitutes a predictor of overall average client's load profile $\mathbf{Y}$. \smallskip\\
\noindent
\setcounter{tblEqCounter}{\theequation} %at the start of the table, set the counter to equation numbering
\begin{table}[ht!]
%\caption{Basic Table}
\label{tab:1}
\centering
\caption{Performances of MECS and MEC in terms of computational complexities (Comp. Time), loss of covariance information (Eigenvalues) and Entropy maximization (Entropy). }
\small
\centering
\begin{tabular}{ccccccc p{5cm}}
\arrayrulecolor{blue}\toprule
\multirow{2}{*}{Data}&\multicolumn{2}{c}{Time (Minutes)}&\multicolumn{2}{c}{Eigenvalues}&\multicolumn{2}{c}{Entropy}   \\\cline{2-7}
%\toprule
& MECS	&MEC &MECS &MEC &MECS &MEC \\ \arrayrulecolor{blue}
\toprule
SMD.  & 10.122  &4.7838 &$\left[0.000,478.625\right]$ &$\left[0.082,107.850\right]$ &1290.965 &647.2493\\
Leukemia.  &61.365   &58.5064 & $\left[0.000,1.306\times 10^{9} \right]$&$\left[6.818\times 10^{5},9.84\times 10^{8} \right]$ &7379340395 &10144024606\\
\bottomrule
\end{tabular}
\end{table}
\setcounter{equation}{\thetblEqCounter} %at the end of the table, set the equation numbering to the counter
\noindent
To compare MECS method with the proposed MEC estimator in terms of addressing loss of covariance information, entropy maximization and computational cost, eigenvalue minimum, eigenvalues maximum and entropy estimates of both covariance estimators were considered. All computations were performed on a personal laptop with the following specifcations: processor Intel(R) Core(TM) i7-7500U CPU \@ 2.70 GHz 2.90GHz 16 GB RAM 64-bit Windows operating system. Table 1 shows the results of estimating the MEC and MEC estimators from microarray and utility billing data sets.
\begin{table}[ht!]
%\caption{Basic Table}
\centering
\caption{Application Performances of MEC with Sliced Inverse Regression (MEC-SIR) and Sliced Average Variance Estimation (MEC-SAVE): The Absolute Correlations between Y and the First Two Sufficient predictors $\left(\left| \mbox{COR}(\mathbf{X}\widehat{\mathbf{\beta}}_{1},\mathbf{Y})\right|,~ \left| \mbox{COR}(\mathbf{X}\widehat{\mathbf{\beta}}_{2},\mathbf{Y})\right|\right) $ and the standard error of estimates.}
\label{tab:2}
%\small
\centering
\begin{tabular}{ccccc p{5cm}}
\arrayrulecolor{blue}\toprule
Method& Data &$\left| \mbox{COR}(\mathbf{X}\widehat{\mathbf{\beta}}_{1},\mathbf{Y})\right|$ &$ \left| \mbox{COR}(\mathbf{X}\widehat{\mathbf{\beta}}_{2},\mathbf{Y})\right|$ & $\mbox{SE}(\mathbf{X}\mathbf{\widehat{\beta}}_{1})$  \\\arrayrulecolor{blue}
\toprule
\multirow{2}{*}{MEC-SIR}& SMD&0.880   & 0.035 &0.004 \\
 &Leukemia &1.000&0.000&$8.614\times 10^{-05}$ \\\arrayrulecolor{blue}
\toprule
\multirow{2}{*}{MEC-SAVE} & SMD  &0.200&	0.146&	0.008 \\
&Leukemia.  &0.488	&0.443	&$8.789\times 10^{-05}$  \\
\arrayrulecolor{blue}
\bottomrule
\end{tabular}
\end{table}
The MEC estimator performed better than the existing MECS estimator in terms of circumventing singularity and instability of estimates since MECS eigenvalues still degenerate to zeros in high-dimension and it does not maximize entropy as much as the proposed MEC estimator except slightly with the SMD. The MEC estimator also achieves much less computational cost than the MECS estimator. 
\noindent
MEC and MECS based SIR and SAVE were applied to the training data to obtain an estimated basis vector $\widehat{\mathbf{\beta}}$. To assess estimation accuracy, two metrics: the absolute correlation between the first two estimated sufficient predictors and the response variable $\left(\left| \mbox{COR}(\mathbf{X}\widehat{\mathbf{\beta}}_{1},\mathbf{Y})\right|,~ \left| \mbox{COR}(\mathbf{X}\widehat{\mathbf{\beta}}_{2},\mathbf{Y})\right|\right), $ and standard error of estimated sufficient predictors from binary logistic regression (leukemia data)  and ordinary least squares regression (Smart meter data) models were computed. The closer the absolute correlation to
one the better estimation of the central subspace. Smaller standard error estimates indicate high quality of model fit.\\
In terms of small standard error of estimates, summaries in table \ref{tab:2} reveal that high accuracies are achieved in both logistic regression classification and least squares regression models with only first MEC-SIR and MEC-SAVE estimated directions as predictors. The first MEC-SIR estimated direction is more predictive of the the response than that of the MEC-SAVE in terms of absolute correlation values. This is expected since SIR is known to recover any vector in the central subspace $\mathbf{S}_{\mathbf{Y}|\mathbf{X}}$ as long as the regression function is not symmetric about 0 \citep[p. 48]{CookWeisberg1991, Li2018}.
\begin{table}[ht!]
%\caption{Basic Table}
\centering
\caption{Statistical Classification Performance of SDR Method with MEC Estimator.}
\label{tab:3}
%\small
\centering
\begin{tabular}{ccccccccc p{5cm}}
\arrayrulecolor{blue}\toprule
Method&	Data&	Classifier&	CCR(\%)&TPR(\%)&FPR(\%)&PPV(\%)&NPV(\%)&AUC(\%) \\\arrayrulecolor{blue}
\toprule
\multirow{6}{*}{MEC-SIR}& \multirow{6}{*}{Leukemia}&LR & 100&100&0.00&100&100&100 \\
 &&LDA&100&100&0.00&100&100&100 \\
 && QDA&100&100&0.00&100&100&100\\
 &&1-NN&100&100&0.00&100&100&100\\
 &&NBayes&100&100&0.00&100&100&100\\
 &&CTree&100&100&0.00&100&100&100\\
 \arrayrulecolor{blue}
\toprule
\multirow{6}{*}{MEC-SAVE}& \multirow{6}{*}{Leukemia}&LR&76&43&0.00&100&71&86\\
&&LDA&76&43&0.00&100&71&86\\
&&QDA&100&100&0.00&100&100&100\\
&&1-NN&100&100&0.00&100&100&100\\
&&NBayes&98&93&0.00&100&95&98\\
&&CTree&85&64&0.00&100&80&90\\
\arrayrulecolor{blue}
\bottomrule
\end{tabular}
\end{table}\\
\noindent
To assess classification performances with first MEC-SIR and MEC-SAVE predictors, we report six metrics: the correct classification rate (CCR) defined as fraction of predictions the method gets right, the true positive rate (TPR) defned as the proportion of truly positive Leukemia outcomes that are identified as positive by the method, the false positive rate (FPR) defined as the proportion of truly negative Leukemia outcome that are identified as positive by the method, the positive predictive value (PPV) defined as how well positive Leukemia outcomes as identified by the model predict actual presence of Leukemia in patients, negative predictive value (NPV) defined as how well negative Leukemia outcomes as identified by the model predict actual absence of Leukemia in patients, and the area under the curve (AUC) defined as area under the curve of plot FPR vs TPR at different points in $[0, 1]$. Greater AUC value indicate better performance the model. TPR, PPV and NPV estimates close to one and FPR estimates close to zero indicate better classification performances.\\
Table \ref{tab:3} presents classification performances of six statistical classifiers based on first MEC-SIR and MEC-SAVE predictors. The MEC estimator is successful with all classifiers with CCR, TPR, PPV and NPV of 100\% and FPR of zero in SIR-MEC. The CCR, TPR, PPV and NPV also hover around 100\% except LR, LDA and CTree in MEC-SAVE. The FPR estimate is also zero in MEC-SAVE. All classifiers achieved AUC of 100\% in MEC-SIR and atleast 86\% in MEC-SAVE. 
\begin{table}[ht!]
%\caption{Basic Table}
\centering
\caption{Ordinary Least Squares Regression Performance with SDR-MEC predictors.}
\label{tab:4}
%\small
\centering
\begin{tabular}{cccccc p{5cm}}
\arrayrulecolor{blue}\toprule
Method&	Data&$\mathbf{X}\mathbf{\widehat{\beta}}_{1}$&	MSE&$\mbox{Adj.}\mathbf{R}^{2}$&	P-value  \\\arrayrulecolor{blue}
\toprule
MEC-SIR&SMD&0.760&0.091&0.689&$1.38\times 10^{-06}$\\\arrayrulecolor{blue}
\toprule
MEC-SAVE&SMD&0.007&0.079&0.113&0.0136  \\
\arrayrulecolor{blue}
\bottomrule
\end{tabular}
\end{table}
\setcounter{equation}{\thetblEqCounter} %at the end of the table, set the equation numbering to the counter\\
\noindent
In the regression application, MEC is also successful with just first MEC-SIR and MEC-SAVE as predictor variable. Small mean square errors, small p-values and adjusted coefficient of determination further confirm this in table \ref{tab:4}. In terms of inference, The estimated first MEC-SIR and MEC-SAVE direction is significant with p-values smaller than 5\%.  
\newpage
\noindent
Figure \ref{fig:1} presents the boxplots of the sufficient predictors for the training (first column) and the test (second column) data from the MEC-SIR and MEC-SAVE estimates. The boxplots in the first row show that MEC-SIR does well in classifying the two groups in both the training and test data. However, the boxplots in the second row reveal that MEC-SAVE does not do as well in classifying the two groups in the test data. 
\newpage
\noindent
ROC curves in figure \ref{fig:2} reveal that there is detectable difference between classification performances of the classifiers with MEC-SIR and MEC-SAVE and random classification performances. This further supports inferences from other reported performance metrics. 
\newpage
\section{Conclusion}
This work has proposed the Maximum Entropy Covariance (MEC) estimator for sufficient dimension reduction with ultrahigh regression and classification problems. The MEC method utilizes the ME principle in a more prudent way in forming convex covariance mixture than the MECS method since it saves a lot of computational time and prevents covariance singularity and instability even in cased where the MECS method fail. It also outperforms the MECS method by preventing the potential errors that may arise in applications where the assumption that all classes (sample groups) have similar covariance shapes may be wrong. Unlike the MECS method which is based on \textquotedblleft selecting the most reliable dispersions of a convex mixture of covariance matrices and thus may not lead to the highest classification accuracy in all circumstances \textquotedblright, the proposed MEC estimator utilizes the most stable and informative convex mixture of covariance matrices to achieve highest classification and regression accuracies in statistical covariance based methods in limited-sample-size problems. The proposed MEC estimator has also been demonstrated in this work to efficiently deal with singularity and instability of sample covariance estimate in SDR applications without requirement for time-consuming covariance estimation procedures. The proposed MEC estimator fully addresses loss of covariance information in ultrahigh regression and classification problems.

\bibliographystyle{Apalike}
\newpage
\bibliography{Bibliography-MM-MC}
\end{document}